\documentclass[pre,twocolumn,superscriptaddress,showpacs]{revtex4}
\usepackage{bm}  
\usepackage{amsmath}
\usepackage{amssymb}
\usepackage{graphicx}
\newcommand{\bra}{{\langle}}
\newcommand{\ket}{{\rangle}}
\newcommand{\be}{\begin{equation}}
\newcommand{\ee}{\end{equation}}
\newcommand{\ba}{\begin{eqnarray}}
\newcommand{\ea}{\end{eqnarray}}
\newcommand{\up}{\uparrow}
\newcommand{\down}{\downarrow}
\newcommand{\nor}[1]{\mathrm{#1}}

\begin{document}
\title{Topological doping of repulsive Hubbard models}

\author{A. Kesting}
\address{Institute for Economics and Traffic, Dresden University
of Technology, Andreas-Schubert-Strasse 23, D-01062 Dresden, Germany}
\email{kesting@vwisb7.vkw.tu-dresden.de}
\affiliation{Institut f{\"u}r Theoretische Physik, Freie Universit{\"a}t
Berlin, Arnimallee 14, D-14195 Berlin, Germany}
\author{C. Timm}
\affiliation{Institut f{\"u}r Theoretische Physik, Freie Universit{\"a}t
Berlin, Arnimallee 14, D-14195 Berlin, Germany}
\email{timm@physik.fu-berlin.de}
\date{August 12, 2003}

\begin{abstract}
The spin configuration induced by single holes and hole pairs doped
into stoichiometric, antiferromagnetic cuprates is considered.
Unrestricted Hartree-Fock calculations for the three-band Hubbard model
are employed to study spin-polaron and vortex-like (meron) solutions.
Meron solutions for a single hole are found to be metastable with higher
energy than spin polarons. We observe that the meron solution shifts from
site-centered to bond-centered as the interaction is increased.
Meron-antimeron solutions for hole pairs are found to be unstable. The
results are in agreement with earlier findings for the one-band Hubbard
model. However, we find that the Hubbard interaction of the one-band
model has to be chosen similar to the one of the three-band model to
obtain comparable results, not of the order of the charge-transfer gap,
as previously expected.
\end{abstract}
\maketitle

\section{Introduction}

Superconductivity appears in cuprates upon introduction of electrons or
holes into the insulating, antiferromagnetic parent compounds. These are
insulators due to the strong electron-electron interactions, which lead
to a Mott-Hubbard splitting of the bands. Upon doping with holes, the
antiferromagnetic order is lost at about 2\% holes per copper atom and
the system becomes superconducting at a somewhat higher doping level.
The doping of Mott-Hubbard insulators has been studied quite
extensively. Nevertheless there is still no agreement on what happens
when a few electrons or holes are introduced.

{}From experiments as well as from band-structure calculations it is known
that the undoped compounds have one hole in each Cu d-shell, whereas
the oxygen-dominated bands are completely filled. Consequently, the Cu
ions carry spins $1/2$, which order antiferromagnetically due to
superexchange. Electron and
hole doping are qualitatively different: Electrons mostly fill the Cu hole
so that the local spin vanishes. Holes predominantly reside in the planar
oxygen orbitals, introducing another spin $1/2$. This physics can be
modelled by the three-band Hubbard Hamiltonian
\begin{equation}\label{1.Hub3}
\begin{split}
\mathcal{H}_\nor{3B}  = &
\: \epsilon_d\sum\limits_{i\sigma} d_{i\sigma}^\dagger\,d_{i\sigma} +
 \epsilon_p\sum\limits_{j\sigma} p_{j\sigma}^\dagger\,p_{j\sigma} + \\
& t_{pd}^\pm \sum\limits_{\bra ij\ket\sigma}  \left( p_{j\sigma}^\dagger d_{i\sigma}+\mathrm{h.c.}\right)+
 U_d\sum\limits_{i} n^d_{i\up} n^d_{i\down},
\end{split}
\end{equation}
where $d_{i\sigma}^\dagger\: (p_{j\sigma}^\dagger)$ is the creation
operator of an electron in a 3d$_{x^2-y^2}$ (2p$_x$, 2p$_y$) orbital at
site $i$ ($j$) and $n^d_{i\sigma}=d_{i\sigma}^\dagger\,d_{i\sigma}$ is
the corresponding density operator.
The parameters $\epsilon_d$, $\epsilon_p$ denote onsite energies.  The
hopping amplitude $t_{pd}^\pm$ between nearest-neighbor copper and
oxygen sites $\bra ij\ket$ changes sign due to the two possibilities
of overlap, \textit{e.g.}, positive (negative) for the bonds pointing
from a Cu site in the positive (negative) $x$ and $y$ directions.
Finally, $U_d$ is the onsite Coulomb interaction for two
electrons on the same Cu site with opposite spins.

The Coulomb interaction $U_p$ between oxygen electrons is irrelevant for
the case of one or two holes, since the probability of two holes sitting
on the same oxygen ion is extremely low. Furthermore, the hybridization
between the oxygen orbitals is omitted as the hopping  $t_{pd}$ is the
dominant process. Furthermore we neglect the Coulomb interaction between
Cu and O electrons.

Electron and hole doping also have rather different experimental
consequences: Antiferromagnetic order is destroyed by only $x\sim 0.02$
holes per Cu, whereas it survives on the electron-doped side up to $x\sim
0.14$. This asymmetry is easily understood: Electrons dilute the
antiferromagnet, which perturbs the order relatively weakly. On the other
hand, holes introduce extra oxygen spins, which couple
antiferromagnetically to both their neighboring Cu spins, resulting in a
strong \emph{ferromagnetic} coupling between those spins \cite{Aha}. This
ferromagnetic bond strongly frustrates the antiferromagnet, leading to its
rapid destruction \cite{Aha,TB,KS}. This picture is supported by
experiments on cuprates doped with nonmagnetic impurities such as zinc
\cite{zinc,Vajk}. They dilute the spin system similarly to electron doping,
leading to a large critical concentration for the destruction of
antiferromagnetic order \cite{Vajk}, which is essentially a percolation
transition \cite{AS}.

The qualitative difference between electron and hole doping suggests that a
small number of charge carriers doped into the parent compounds should also
behave quite differently. Indeed, several authors propose that hole doping
leads to the appearance of nontrivial structures of the surrounding Cu
spins, such as dipolar configurations \cite{Aha,Shrai}, vortices or merons
\cite{Verges,Sei98,BJ}, skyrmions \cite{Wieg,Good,Mari99}, and domain walls
or stripes \cite{Verges,Tranq,EKstripe,Sei98,WS}. This should be contrasted
with a spin-polaron configuration, where the spins in the vicinity of the
hole might be reduced or even inverted in sign but are still collinear.
Timm and Bennemann \cite{TB} have proposed a mechanism for the rapid
destruction of antiferromagnetic order in the hole-doped regime that relies
on the formation of merons in the staggered magnetization upon doping. We
here use the term ``meron'' to refer to a topological defect in the
easy-plane Heisenberg model. It is very similar to a vortex in the
\textit{XY} model, but the order parameter may rotate out of the easy plane
close to the defect center in a ``lotus flower'' configuration. The
easy-plane anisotropy is caused by the Dzyaloshinskii-Moriya interaction
\cite{DM,Coffey}, resulting from buckling of the CuO$_2$ planes and
spin-orbit coupling. For energy reasons, merons are expected to be created
as meron-antimeron pairs. It is suggested that merons appear since each
hole introduces a ferromagnetic Cu-Cu bond, leading to frustration, which
can be reduced by forming merons centered on each such bond \cite{TB}. This
mechanism only works for hole doping, not for electron doping, in agreement
with the pronounced asymmetry of the phase diagram.

The question of the existence of meron-antimeron pairs in the ground
state has been studied with various approaches \cite{Verges,Sei98,BJ}.
The present situation is not satisfactory, since unrestricted
Hartree-Fock calculations for the one-band Hubbard model have found
meron-antimeron pair solutions, but never as the ground
state \cite{Verges}. On the other hand, Hartree-Fock calculations have
been performed for the three-band model \cite{Verges3,LopS}, but
meron-antimeron pairs have not been considered.

The one-band
Hubbard model can be derived from the three-band model, Eq.~(\ref{1.Hub3}),
as an effective low-energy model \cite{ZhangRice}.
The one-band model is defined by the Hamiltonian
\be
\mathcal{H}_\nor{1B} =
  - t \sum\limits_{\bra ij\ket,\sigma} \big( c_{i\sigma}^\dagger
    c_{j\sigma} + \mathrm{h.c.} \big)
  + U \sum\limits_{i} c^\dagger_{i\up}c_{i\up} c^\dagger_{i\down}
    c_{i\down} .
\ee
In the undoped three-band model the Fermi energy lies between an
occupied oxygen-type band and the unoccupied upper Hubbard band.
The charge-transfer gap
between them is mimicked by the Mott-Hubbard gap of the one-band model
and the onsite interaction $U$ of the one-band model must be
chosen accordingly \cite{ZSA,Brenig}.
However, this derivation \cite{ZhangRice} of the
one-band model assumes a polaron-like configuration for a doped hole and
would not work if meron-antimeron pairs were the true ground state.

On the other hand, a slave-boson approach did find meron-antimeron pairs
as the ground state for two holes, again in the one-band
model \cite{Sei98}. This approach goes beyond the Hartree-Fock
approximation (HFA) and leads to lower variational ground-state
energies. Furthermore, in a one-band spin-flux model Berciu and
John \cite{BJ} also find meron-antimeron pairs as the ground state,
employing the configuration-interaction (CI) method, which also goes
beyond Hartree-Fock theory. Both approaches have not yet been applied to
the three-band model. The question appears to be open.

In the present paper we consider the three-band Hubbard model within the
HFA. In Sec.~\ref{sec.th} we briefly outline the theory. In
Sec.~\ref{sec.res}
we compare the results with Hartree-Fock calculations for the
one-band model for larger system sizes than studied previously. We study
the stability of spin-plasmon and meron solutions for a single hole and,
in particular, of meron-antimeron solutions for two holes. We also
consider changes in the meron structure as a function of the interaction
strength.

\section{Theory}
\label{sec.th}

Since it is mostly standard, we can be brief in describing our
implementation of the unrestricted Hartree-Fock approximation
\cite{diplom}. We here
do this for the three-band Hubbard model. The HFA reduces the many-body
problem to a set of single-electron problems by decoupling the interaction,
in our case in the particle-hole channel.
To include transverse spin degrees of freedom we decouple on the 
level of creation and annihilation operators,
\begin{equation}
\begin{split}
& U_d \sum_{i} d_{i\up}^\dagger d_{i\up} \,d_{i\down}^\dagger d_{i\down}
  \\
& \cong U_d \sum_{i}  \left(
  d_{i\up}^\dagger, d_{i\down}^\dagger\right)
\left(\begin{array}{cc}
\bra n_{i\down}^d\ket  & -\bra S_{i}^-\ket\\
-\bra S_{i}^+\ket  & \bra n_{i\up}^d\ket
\end{array}\right)
\left(\begin{array}{c}
d_{i\up}\\
d_{i\down}
\end{array}\right) -E_0,
\end{split}
\end{equation}
with the spin-flip operators $S^+=S^x+iS^y=d_\up^\dagger d_\down$ and
$S^- = S^x-iS^y = d_\down^\dagger d_\up$ and
the constant energy shift
\begin{equation}
E_0 = U_d \sum_{i} \left(\frac{1}{4} \bra \,n_{i}^d
    \ket^2 - \bra  S_{i}^z \ket^2-\bra  S_{i}^x \ket^2
    - \bra S_{i}^y \ket^2\right).
\end{equation}
After decoupling the interaction, the many-particle ground state
$\vert\Psi\ket$ for $T=0$ in the HFA is constructed by creating $N_e$
electrons in the lowest-energy single-particle states out of
the vacuum $\vert 0\ket$,
\begin{equation}
\vert \Psi\,\ket =\prod_{n=1}^{N_e} a_n^\dagger\, \vert 0\,\ket .
\end{equation}
The single-particle states are given in terms of
wave functions $\phi_n$ by
\begin{equation}
a_n^\dagger =  \sum_{i\sigma} \phi_n(i,\sigma)\,c_{i\sigma}^\dagger,
\end{equation}
where the sum over $i$ runs over all relevant
Cu d$_{x^2-y^2}$ and oxygen p$_x$ and p$_y$ orbitals,
and $c_{i\sigma}^\dagger$ creates an electron in orbital $i$.

The Hartree-Fock eigenequations are derived by minimizing the
expectation value to the total energy in the state
$\vert \Psi\ket$ with the $\phi_n(i,\sigma)$ as variational parameters.
For the $n$-th state with spin $\sigma=\uparrow, \downarrow$
at the Cu site $i$ the equation is of the form
\begin{equation}
\label{hf_eigenequations}
\begin{split}
 &E_n\,  \phi_n(i,\sigma) = -t_{pd}^\pm \sum_{j\in V_i} \phi_n(j,\sigma)\,+ \\
& \sum_\nu \left(\left(\frac{U_d}{2}+\epsilon_d\right) \bra n_{i}^d\ket
  \delta_{\sigma\nu}
  - U_d \bm{\sigma}_{\sigma\nu}\bra \mathbf{S}_i\ket\right) \phi_n (i,\nu).
\end{split}
\end{equation}
The notation $j\in V_i$ means that the sum is per\-formed over the sites
$j$ which are nearest neighbors of site $i$.
$\bm{\sigma} = (\sigma^x, \sigma^y, \sigma^z)$ is the vector of Pauli
matrices and $\sigma^0$ is the unit matrix.
The local Hartree-Fock fields of the charge and spin density,
\begin{eqnarray}
\bra n_i^d\ket & = & \bra\Psi\vert \sum_{\mu\nu} d_{i\mu}^{\dagger}
  \sigma^0_{\mu\nu} d_{i\nu} \vert \Psi \ket = \sum_{n\sigma}
  \vert\phi_n(i,\sigma)\vert^2,\\
\bra \mathbf{S}_i \ket &= & \frac{1}{2}\bra \Psi\vert
  \sum_{\mu\nu} d_{i\mu}^{\dagger} \bm{\sigma}_{\mu\nu} d_{i\nu}
  \vert \Psi \ket  \nonumber \\
& = & \frac{1}{2}\sum_{n\mu\nu} \phi_n^*(i,\mu)
  \bm{\sigma}_{\mu\nu} \phi_n(i,\nu)
\end{eqnarray}
are computed self-consistently. The sum over $n$ is performed
for the $N_e$ electrons in the system. The terms for the oxygen sites $j$
are of an analogous form. Finally, the Hartree-Fock
ground-state energy is given by the sum of the eigenvalues $E_n$ for
occupied states,
\begin{equation}
E_N = \sum_{n=1}^{N_e} E_n - E_0.
\end{equation}
We solve the Hartree-Fock equations (\ref{hf_eigenequations}) by
iteration, which requires a numerical diagonalization at each step.
Self-consistency is assumed when the difference in all spin and charge
components is less than $\epsilon\simeq 10^{-5}$.
Typically, the charge distribution converges very rapidly to
the self-consistent value, while the spin configuration needs a lot more
iteration steps. The number of steps depends strongly on the
initial configuration and the boundary conditions.

\section{Results and discussion}
\label{sec.res}

In the present section we examine the Hartree-Fock solutions for the
three-band Hubbard model for one and two holes \cite{diplom}.
The results are compared
with corresponding ones for the one-band model. The calculations have
been carried out for a square lattice with up to $14\times 14$ unit
cells for the three-band model as well as for the one-band model. We have
treated periodic as well as open boundary conditions (BC). The latter
disturb the elementary cells at the edge of the lattice in the
three-band case. We have studied edges of alternating copper and oxygen
ions but also edges made up entirely of oxygen ions.
Finite size effects
are stronger for open BC, as expected, but are neglectible for the lattice
sizes used here. Relevant parameters in the three-band model are (i) the
hopping $t_{pd}$, (ii) the Coulomb interaction $U_d$ for copper sites
and (iii) the band gap $\Delta=\epsilon_p -\epsilon_d$ between the
oxygen and copper orbitals. A Coulomb interaction between oxygen
electrons and a non-zero oxygen-oxygen hopping amplitude do not change
the solutions qualitatively and are neglected in the present study.
A band gap $\Delta>0$ in
combination with $U_d>\Delta$ implies that holes preferentially occupy
oxygen orbitals as observed in experiments. In this regime the occupied
oxygen states and the empty copper states in the upper Hubbard band are
separated by a charge-transfer gap, which is given by $U_d - \Delta$.
For the calculations we have used the parameters taken from
Ref.~\cite{schmalian},
\begin{equation}
\label{used_parameters}
t_{pd}=1\:\nor{eV}, \quad U_d=8\:\nor{eV},\quad \Delta=5\:\nor{eV}.
\end{equation}
The results for the three-band model are compared to corresponding
Hartree-Fock results for the one-band Hubbard model \cite{Verges}
with $t = 1\:\nor{eV}$ and $U$ in the range from $5$ to $8\:\nor{eV}$.

\begin{figure}[htb]
\centering
\includegraphics[width=0.8\columnwidth]{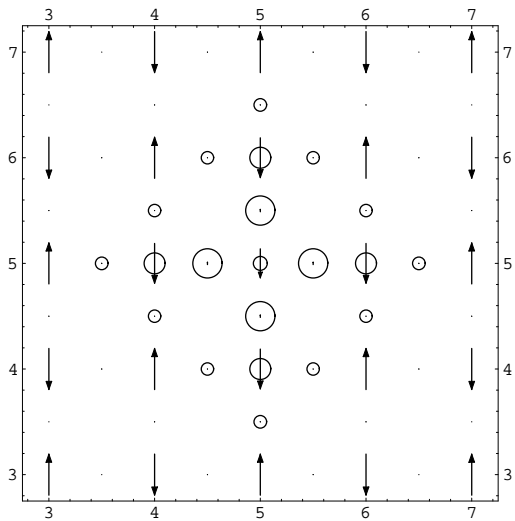}
\caption{Self-consistent Hartree-Fock solution of the three-band Hubbard
model for one hole. The figure shows part of a $10\times 10$
lattice with the spin and relative charge density indicated for each
atomic site by the arrows and circles, respectively.
The reduced copper spin at point $(5,5)$ forms a spin polaron with its four
nearest copper neighbors. The charge is mainly distributed over the four
oxygen neighbors.}
\label{fig1_polaron}
\end{figure}

\subsection{Single-hole solutions}

For a single hole we find a spin-polaron configuration as the
self-consistent solution of the HFA for open as well as periodic
boundary conditions, see Fig.~\ref{fig1_polaron}. In its center a reduced
copper spin (of modulus $0.55$ in units where
the fully polarized state corresponds to unity vs.\ an average
order parameter of $0.78$) is pointing in the same
direction as its four nearest copper neighbors, \textit{e.g.}, a
ferromagnetic ``microdomain'' is formed. The charge is mainly
distributed over the four oxygen neighbors of the central copper ion.
Small magnetic moments at the oxygen sites are induced, which are
oriented oppositely to the copper spins. This result is consistent with
the over-compensation of the antiferromagnetic superexchange by an
effective ferromagnetic Cu-Cu coupling for all values of $U_d > \Delta$,
supporting the picture of Aharony \textit{et al.} \cite{Aha}.

\begin{figure}[htb]
\centering
\includegraphics[width=\columnwidth]{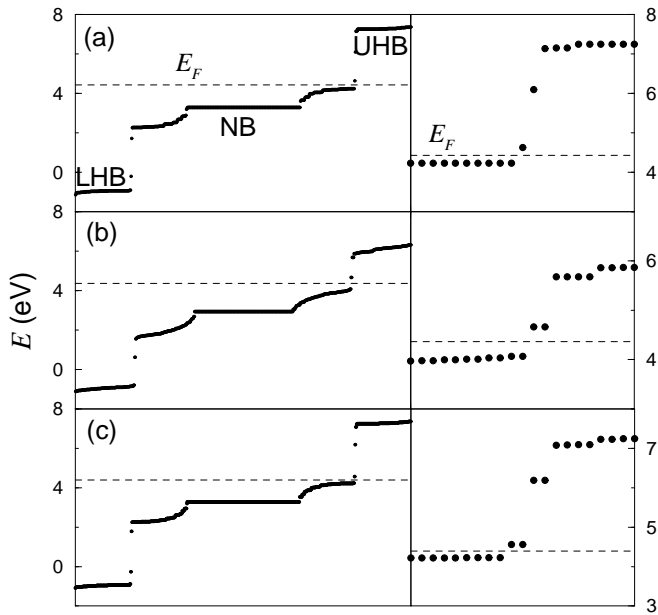}
\caption{One-particle energy spectra for self-consistent solutions of
the HFA. LHB and NB denote the occupied lower Hubbard and non-bonding
bands, respectively. UHB denotes the unoccupied upper Hubbard band
\cite{ZSA,Brenig}. (a) Spectrum for
the spin-polaron solution of Fig.~\ref{fig1_polaron}. (b) Spectrum
for the meron solution of Fig.~\ref{fig2_meron}(a). (c)
Spectrum for the solution of
Fig.~\ref{fig5_solution3band}. The right-hand-side figures are
enlargements of the regions close to the Fermi energy.}
\label{fig6_spectrum}
\end{figure}

In the one-particle spectrum of Fig.~\ref{fig6_spectrum}(a) \emph{two}
states appear in the gap due to the interaction. One is split off from
the occupied oxygen-dominated band but is now empty due to the extra
hole. This is the Zhang-Rice singlet \cite{ZhangRice}. The other state
originates from the unoccupied upper Hubbard band. The first unoccupied
state is to about 75\% localized on oxygen sites emphasizing the oxygen
character of doped holes \cite{91}.

It is worth considering the limit of strong interaction $U_d/t_{pd}\gg
1$ studied by Zhang and Rice \cite{ZhangRice}. For $U_d=50\:\nor{eV}$,
$\Delta =25\:\nor{eV}$ we find a polaron solution similar to
Fig.~\ref{fig1_polaron}, but with the copper
spin moment in the polaron center hardly reduced (to $0.99$) and
the charge mainly (to 80\%) localized on its four
oxygen neighbors. These four ions together carry a spin moment of $-0.79$.
Thus in the HFA the central copper spin forms a singlet
with its four oxygen neighbors, as predicted \cite{ZhangRice}.

It is instructive to compare the results with the one-band Hubbard model,
for which the polaron configuration is also a self-consistent solution in
the HFA \cite{Verges}. For an intermediate interaction of $U\approx 5
\ldots 8\:\nor{eV}$ the charge distribution and the spin moment at the
polaron center are of the same order of magnitude as in the three-band case
so that in this respect the one-band model indeed mimicks the three-band
physics. However, it is surprising that we find this relation for values of
$U$ for the one-band model similar to $U_d = 8\:\nor{eV}$ and \emph{not} to
the charge-transfer gap $U_d-\Delta = 3\:\nor{eV}$, as we would have
expected \cite{ZSA,Brenig}. On the contratry, for $U \lesssim 3\:\nor{eV}$
we find the diagonal cigar-shaped configuration predicted in
Refs.~\cite{SWZ,SC} without any inverted Cu spin \cite{diplom}.
We suggest that $U$ is of the order of
$U_d$ since the local moment is dominated by the interaction in the
copper d orbital and is only marginally affected by the hybridization
with oxygen orbitals.

\begin{figure}[htb]
\centering
\includegraphics[width=0.6\columnwidth]{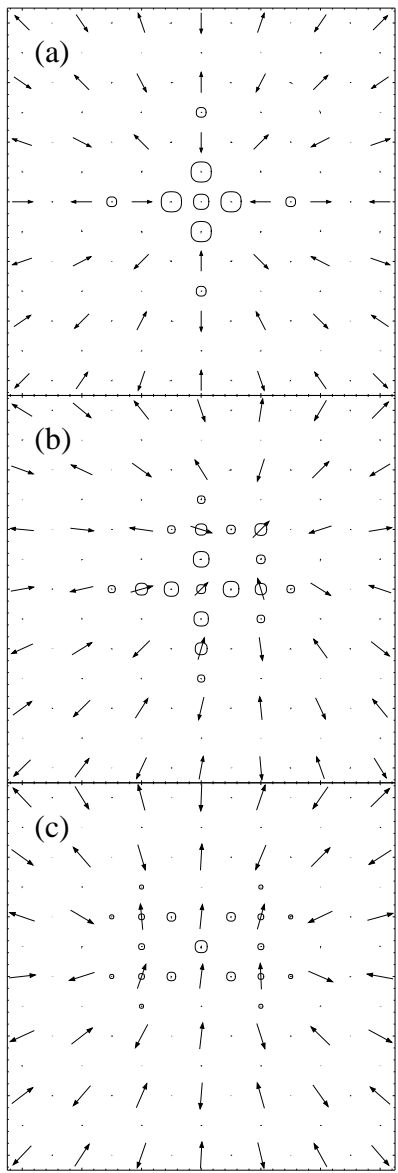}
\caption{Selfconsistent Hartree-Fock solutions with (a)
$U_d=7\:\nor{eV}$, (b)
$U_d=8\:\nor{eV}$ and (c) $U_d=10\:\nor{eV}$ on an $11\times 11$ lattice.
For $U_d\le 7\:\nor{eV}$ site-centered merons are stable solutions
of the HFA.
Increasing $U_d$ shifts the center of meron but preserve the winding
number of the initial configuration.}
\label{fig2_meron}
\end{figure}

\begin{figure}[htb]
\centering
\includegraphics[width=\columnwidth]{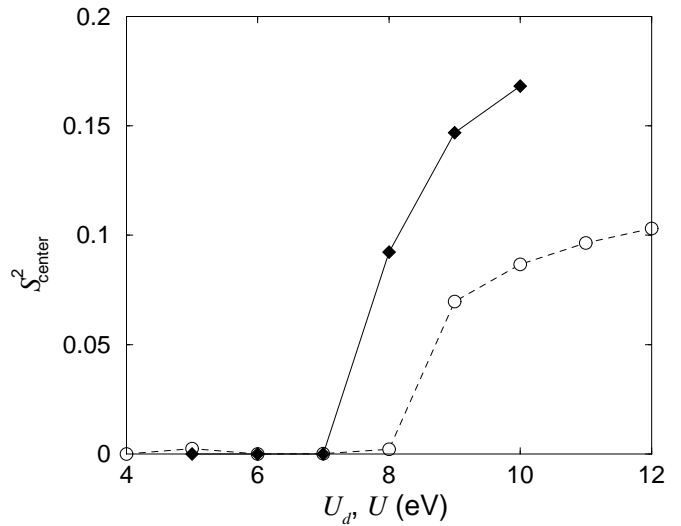}
\caption{The average value $\bra \mathbf{S}\ket^2$ at the site closest to
the meron center
as a function of the interaction $U$ for the one-band model and $U_d$ for
the three-band model. For a site-centered meron, $\bra
\mathbf{S}\ket^2$ is negligible due to symmetry. A non-vanishing value
corresponds to a qualitative change in the spin texture, see
Fig.\protect\ref{fig2_meron}.}
\label{fig3_s2}
\end{figure}

We now turn to meron configurations for a single doped
hole. We start the HFA with a coplanar vortex
configuration. Site-centered Merons turn out to be stable Hartree-Fock
solutions for $U_d \lesssim 7\:\nor{eV}$.
Because of the non-vanishing winding number only
open boundary conditions are possible. In Fig.~\ref{fig2_meron}(a) a
site-centered meron solution for $U_d=7\:\nor{eV}$ is shown
for a lattice with edges of alternating copper and oxygen ions.
But there is no qualitative difference to lattices with edges made up
entirely of oxygen ions.
Increasing $U_d$ shifts the center of the meron first in the direction of
a plaquette and then onto a bond, see Figs.~\ref{fig2_meron}(b) and (c).
In addition, a tendency towards
ferromagnetic alignment, \textit{i.e.}, formation of a spin polaron,
is observed. These results correspond to the one-band solutions for
$U\gtrsim 8\:\mathrm{eV}$, in which
the spin in the meron center is no longer suppressed and shows a tendency
towards formation of a ferromagnetic microdomain.
Note that again $U$ has to be chosen similar to $U_d$ and not to the
charge-transfer gap to obtain this result.
To exhibit the transition from site- to bond-centered merons more
quantitatively, we plot in
Fig.~\ref{fig3_s2} the average $\bra \mathbf{S}\ket^2$
at the site $i=(6,6)$ for the three-band
solution and the corresponding quantity for the one-band model.
For a site-centered meron this number is negligible due to symmetry.

For the one-band model we have compared the energies of polaron and
meron solutions for a single hole by considering their respective
formation energies \cite{excitation_energy}.
As expected, the meron energy is higher than
the energy of the spin polaron and diverges logarithmically with system
size. However, the iteration process of the HFA turns out to be unable
to change the winding number. Adding small random numbers during the
iteration process makes it converge to a spin-polaron configuration.
Thus a meron is a \emph{metastable} solution.

In the one-particle spectrum for the meron solution with
$U_d=7\:\nor{eV}$, Fig.~\ref{fig6_spectrum}(b), two discrete,
degenerate hole states appear: one from the filled oxygen-do\-mi\-nat\-ed
non-bonding band and the other from the unoccupied upper Hubbard-band.
The states are degenerate because an additional electron can be
introduced either with spin up or spin down. The same holds for the
one-band model. A mirror image of these degenerate states appears in
the gap between the occupied bands.

\begin{figure}[htb]
\centering
\begin{tabular}{cc}
\includegraphics[width=0.8\columnwidth]{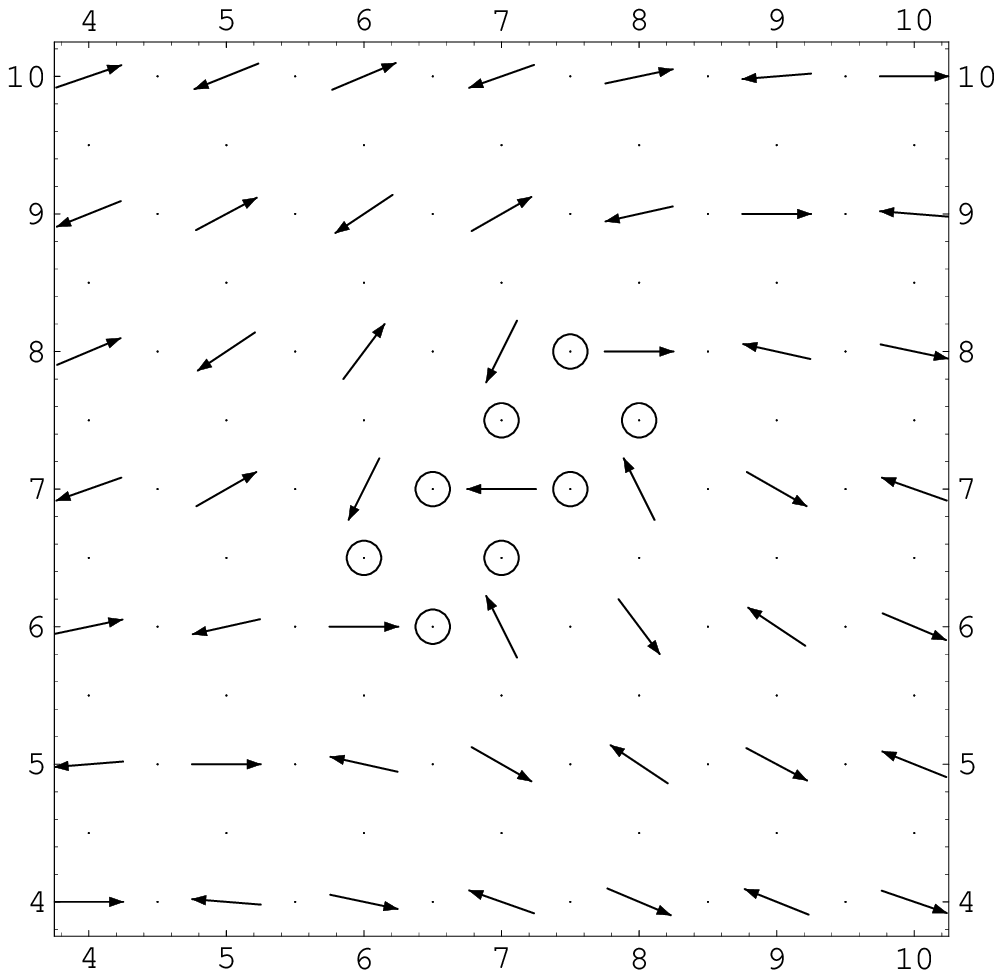} \\
\includegraphics[width=0.8\columnwidth]{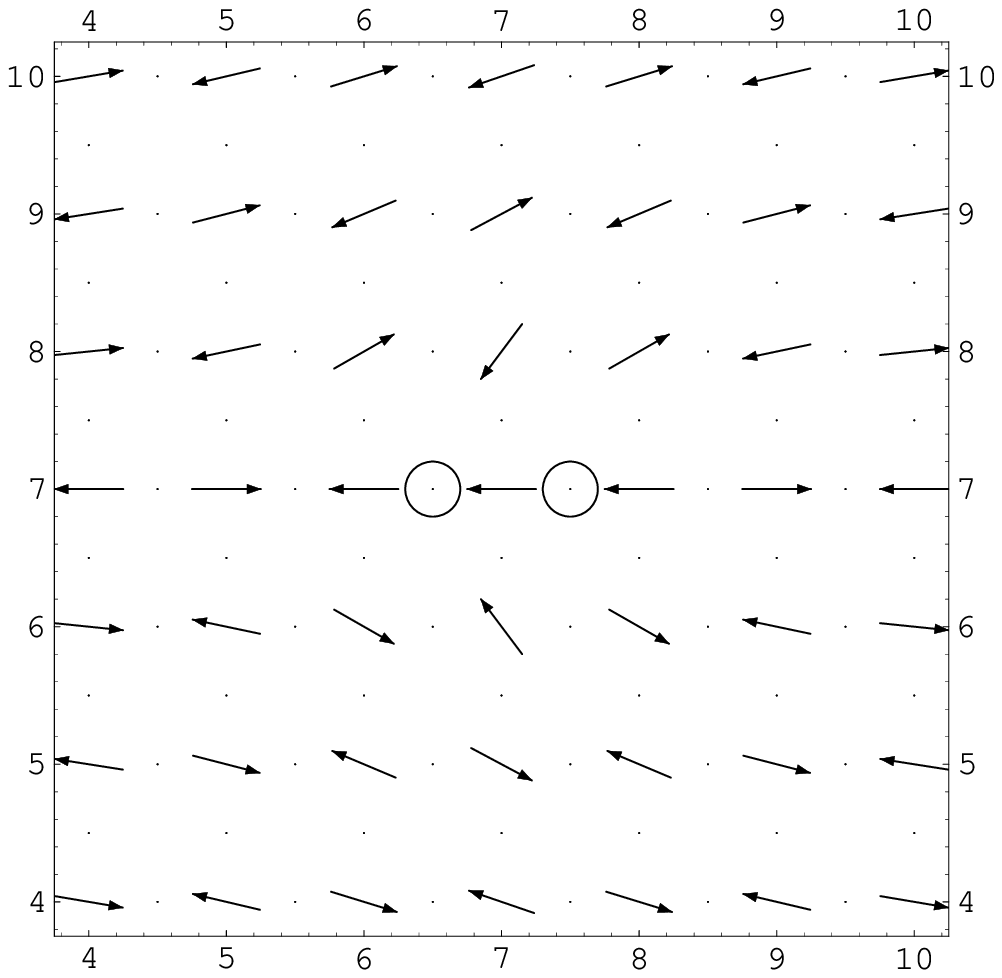} \\
\end{tabular}
\caption{Starting configurations for the spin and charge density for a
system with two holes on a $14\times 14$ lattice: (top)
plaquette-centered meron-antimeron pair and (bottom) bond-centered
meron-antimeron pair.}
\label{fig4_initialconfiguration}
\end{figure}

\begin{figure}[htb]
\centering
\includegraphics[width=0.8\columnwidth]{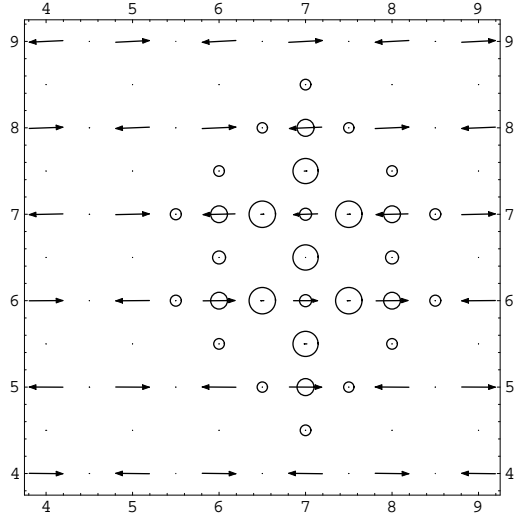}
\caption{Resulting double-polaron configuration for two holes on a
$14\times 14$ lattice with periodic boundary conditions using the
parameters in Eq.~(\ref{used_parameters}).}
\label{fig5_solution3band}
\end{figure}

\subsection{Two-holes solutions}

We now turn to a system doped with \emph{two} holes to investigate
possible meron-antimeron solutions. As ini\-tial con\-figu\-ra\-tions
 we con\-sider pairs of bond-\-cen\-ter\-ed as well as 
plaquette-centered defects, 
see Fig.~\ref{fig4_initialconfiguration}, and both open and periodic BC.
We find that these configurations are not stable under the HFA
iteration. This result holds for all parameter sets we have
investigated. Instead, the system develops into a polaron-type spin
configuration, the detailed structure of which depends on the choice of
BC. For example,
Fig.~\ref{fig5_solution3band} shows the solution for the bond-centered
initial configuration and periodic BC on a $14\times 14$ lattice. We
find two spin polarons with antiferromagnetically aligned center spins.
The polaron centers are localized on neighboring copper sites. At the
oxygen sites small magnetic moments are induced which are oriented in
antiparallel to the copper spins.

The origin of the instability of meron-antimeron solutions is their
higher energy and their vanishing winding number. The latter
means that they can be destroyed during the HFA iteration since no high
energy barrier has to be overcome. Since for large separations
$\mathbf{R}$ of meron and antimeron the interaction increases
logarithmically with $\mathbf{R}$, the energy of the meron-antimeron
solution will eventually become larger than the spin-polaron energy.
However, our results indicate that this is also the case at the smallest
separations.

Again, similar results are obtained for the one-band model.
Bond-centered as well as plaquette-centered meron-antimeron pairs as
initial configurations develop into spin polaron solutions as shown
earlier by Verg{\'e}s \emph{et al.} \cite{Verges}. In contrast,
Seibold \cite{Sei98} has found a stable solution in the one-band model
using a slave-boson approximation beyond Hartree-Fock.

In Fig.~\ref{fig6_spectrum}(c) the spectrum of the discussed
double-polaron solution is shown. For each doped hole one state
splits off from the occupied non-bonding band and another from the
unoccupied upper Hubbard band. The first two hole states are localized
to about 80\% on oxygen sites. Further, the states are
energetically degenerate reflecting the symmetry of the solution under
inversion in the center of the bond between $(7,6)$ and $(7,7)$ in
Fig.~\ref{fig5_solution3band} and simultaneous spin inversion.

\section{Conclusions}

In this paper we have used unconstrained Hartree-Fock theory at $T=0$,
keeping all local spin and charge degrees of freedom to study
inhomogenous spin textures in the three-band and the one-band Hubbard
model. This approach neglects flucutations around the mean-field
solutions. Thus one has to consider the reliability of the these
solutions.  The HFA predicts the correct behavior for large $U_d$ in the
half-filled case. For one doped hole spin polarons are obtained as
solutions for the three-band Hubbard model.  This agrees with the case
of the one-band model \cite{Verges}, in agreement with the mapping
\cite{ZhangRice,Brenig} of the three-band on the one-band model at low
energies. However, we find that in the one-band model an on-site
interaction $U$ of the order of the three-band interaction $U_d$ is
required to obtain similar results.

A meron is only a \emph{metastable} self-consistent solution due to its
nonzero winding number. As our main result we find that meron-antimeron
pairs are not stable solutions in the HFA of the three-band Hubbard
model. This holds for periodic as well as open boundary conditions. The
same is true for the one-band model. Thus a consistent picture of the
one- and three-band models emerges, at least in the framework of the
HFA: Both show the same qualitative behavior for weak hole doping.

While the mean-field approximation serves as an useful tool for studying
the local charge and spin density, it yields only an upper bound for the
true ground-state energy. In the one-band Hubbard model Seibold
\cite{Sei98} found a meron-antimeron ground state for $T=0$ for suitable
parameters using a slave-boson method. This method uses a Hartree-Fock
ansatz as the initial configuration but treats the spin and charge
excitations at each site as distinct bosonic degrees of freedom. It
leads to a lower ground state energy than the HFA and is thus the
superior variational method. We suggest that this method applied to the
repulsive three-band Hubbard model could clarify the still open question
of stable meron-antimeron pairs as the effect of hole doping.

\thanks{We would like to thank G. Seibold and F. Sch{\"a}fer for useful
discussions and comments. A.K. acknowledges support by the Evangelisches
Studienwerk (scholarship organisation of the Protestant Churches in
Germany).}


\begin{thebibliography}{99}
\bibitem{excitation_energy}This is the energy needed to add a hole to the
half-filled system. First, one has to lift an
electron to the Fermi-level which is given by the half of the Mott-Hubbard
gap. A second contribution comes from the change of the configuration energy
due to the interaction of the particles in the system.
\end{thebibliography}
\end{document}